\documentclass{osa-article}

\journal{osajournal}

\articletype{Research Article}

\begin{document}

\title{Accelerating Silicon Photonic Parameter Extraction using Artificial Neural Networks}

\author{Alec M. Hammond,\authormark{1} Easton Potokar,\authormark{1} and \\ Ryan M. Camacho\authormark{1,*}}

\address{\authormark{1}Electrical \& Computer Engineering Department, Brigham Young University, Provo, UT, USA\\}

\email{\authormark{*}camacho@byu.edu} 


\begin{abstract*}
We present a novel silicon photonic parameter extraction tool that uses artificial neural networks. While other parameter extraction methods are restricted to relatively simple devices whose responses are easily modeled by analytic transfer functions, this method is capable of extracting parameters for any device with a discrete number of design parameters. To validate the method, we design and fabricate integrated chirped Bragg gratings. We then estimate the actual device parameters by iteratively fitting the simultaneously measured group delay and reflection profiles to the artificial neural network output. The method is  fast, accurate, and capable of modeling the complicated chirping and index contrast.
\end{abstract*}

\section{Introduction}
Interest in integrated optics continues to grow as silicon photonics provides an affordable platform for areas like telecommunications, quantum information processing, and biosensing \cite{chrostowski_silicon_2015}. Silicon photonic devices typically contain features with sub-micron dimensions, owing to the platform's high index contrast and years of complementary metal-oxide-semiconductor (CMOS) process refinement. \par

While small features enable several innovative and scalable designs, they also induce an increased sensitivity to fabrication defects \cite{bogaerts_silicon_2018}. A fabriction defect of just one nanometer, for example, can cause a nanometer shift in the output spectrum of the silicon photonic device \cite{selvaraja_wafer-scale_2011}. Understanding and characterizing these process defects is essential for device modeling and variability analysis \cite{lu_performance_2017,zortman_silicon_2010}. Predicting and compensating for such sensitivity in the optical domain is difficult because typical simulation routines are computationally expensive and in many cases, prohibitive.

To overcome these challenges, we propose a new parameter extraction method using artificial neural networks (ANN). We train an ANN to model the complex relationships between integrated chirped Bragg gratings (ICBG) \cite{wang_narrow-band_2012,strain_design_2010} and their corresponding reflection and group delay profiles. We use the trained ANN to extract the physical parameters of various fabricated ICBGs using a nonlinear least squares fitting algorithm --- a task that is computationally prohibitive using traditional simulation routines. We find that the proposed routine produces spectra that matches well the experimental reflection and group delay profiles for the ICBGs.

Our work builds upon previous efforts that extract integrated photonic device parameters using analytic models. Chrostowski \textit{et al}., for example, extract the group index across a wafer with 371 identical microring resonators (MRR) using an analytic formula describing the free spectral range (FSR) \cite{chrostowski_impact_2014}. Similarly, Chen \textit{et al}. derive both the effective and group indices from MRRs by fitting the full, analytic spectral transfer function to the experimental data \cite{chen_parameter_2014}. Melati \textit{et al}. extract phase and group index information from small lumped reflectors knows as point reflector optical waveguides (PROW) using various analytic formulas \cite{melati_waveguide-based_2016}. \par

Perhaps most similar to this work, Xing et al. build a regression model from data generated by an eigenmode solver that relates waveguide design parameters (e.g. width and thickness) to their corresponding effective indices \cite{xing_accurate_2018}. They subsequently use this regression model in addition to an analytic transfer matrix and a fitting routine to extract the average waveguide width and thickness from various Mach-Zhender Interferometer (MZI) devices. Our ANN parameter extraction method is fast, just like the analytic and regression models, but capable of modeling much more complicated devices, like ICBGs.

The rest of the paper is outlined as follows: first, we describe the data generation process necessary to train the ANN. Next, we describe the process of training the ANN. We then discuss the ICBG device design, fabrication, testing, and data calibration. Finally, we describe our nonlinear fitting algorithm and present our experimental results.

\section{Data generation}

ANNs model the relationship between inputs and outputs by cascading various nonlinear computational units known as neurons \cite{hornik_approximation_1991}. ANN training algorithms like backpropogation tune the weights of these neurons until the functional mapping adequately models the corresponding training set \cite{cun_theoretical_1988}. Several factors, like the ANN's architecture and even the training set itself, influence the training accuracy and speed of the ANN. In addition, the ANN may learn unintentional biases if the training set insufficiently represents the function space \cite{schmidhuber_deep_2015}. \par

Consequently, it is important to adequately describe the ICBG using parameters that are simple and intuitive for the designer, but also comprehensive and descriptive in order to fully span the design space. To accomplish this, we parameterized the ICBG's design space using the length of the first ICBG period ($a_0$), the length of the last ICBG period ($a_1$), the number of gratings ($NG$), and the ICBG's corrugation width ($\Delta w=w_1 - w_0$), and the wavelength ($\lambda$). Figure \ref{fig:processFlow} illustrates an ICBG with each of these design parameters. We trained the ANN to output the reflection and group delay spectra of the simulated ICBG. \par

To accurately simulate the the ICBG reflection and group delay responses for our training-set, we used a layered-dielectric media transfer matrix method (LDMTMM) \cite{helan_comparison_2006}. The method discretizes the ICBG into individual dielectric slabs, models each slab as an ideal waveguide, and propogates the fields through each slab using a transfer matrix. The effective index of each section is modeled using another ANN that parameterizes the wavelength as a function of waveguide width and thickness. This process is repeated for every wavelength point of interest. \par

We simulated over 100,000 grating configurations at 250 wavelength points from 1.45 $\mu$m to 1.65 $\mu$m resulting in approximately 26,000,000 training points. We swept through 10 different corrugation widths, 11 different ICBG lengths, and 961 different chirping patterns \cite{hammond_designing_2018}. \par

For a tool intended to perform parameter extraction on fabricated devices, it is important to use a generalized and abstracted model insensitive to minor fabrication defects. For example, small changes in the ICBG apodization profile greatly alter the expected spectral ringing. The LDMTMM method also exhibits significant ringing that corresponds to a very narrow parameter space. Furthermore, parameterized high frequency ringing is rather difficult to capture efficiently using ANNs. We overcame these challenges by fitting the LDMTMM group delay and reflection profiles prior to training to a generalized skewed Guassian function of the form
\begin{equation}
f(\lambda,\lambda_0,\sigma,\beta,a,p,c) = \frac{a \sigma}{\gamma} e ^{\frac{-\beta|\lambda - \lambda_0|}{\gamma} ^ p} + c
\end{equation}
where 
\begin{equation}
\gamma = \frac{2\sigma}{1+e^{-\beta (\lambda-\lambda_0)}}
\end{equation}
Our resultant dataset corresponds to a much larger and practical parameter space but significantly alleviates the ANN training process. \par

\begin{figure}
	\centering
	\includegraphics{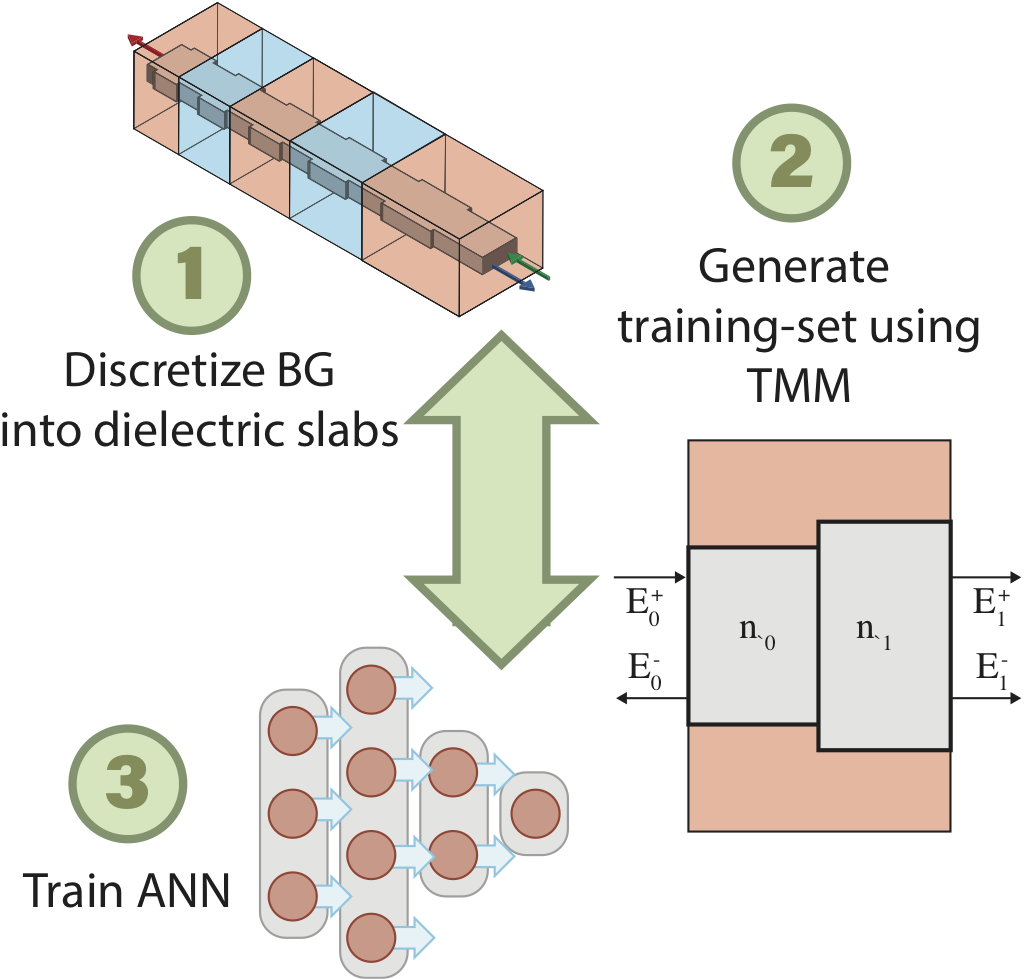}
	\caption{ANN modeling process. First, several different ICBGs are discretized into individual dielectric layers (1). Then, the reflection and group delay profiles are simulated using the transfer matrix method (2). This dataset is then fed into a ANN training algorithm (3). Often, this process must be repeated until the ANN can suitably express a large enough ICBG design space.}
	\label{fig:ANN}
\end{figure}

Once the dataset was generated and processed, we proceeded to train the ANN. Often, this process must be repeated until a suitable parameter space is simulated. This design flow is illustrated in Figure \ref{fig:ANN}.

\section{ANN training}

\begin{figure}
	\centering
	\includegraphics[width=5in]{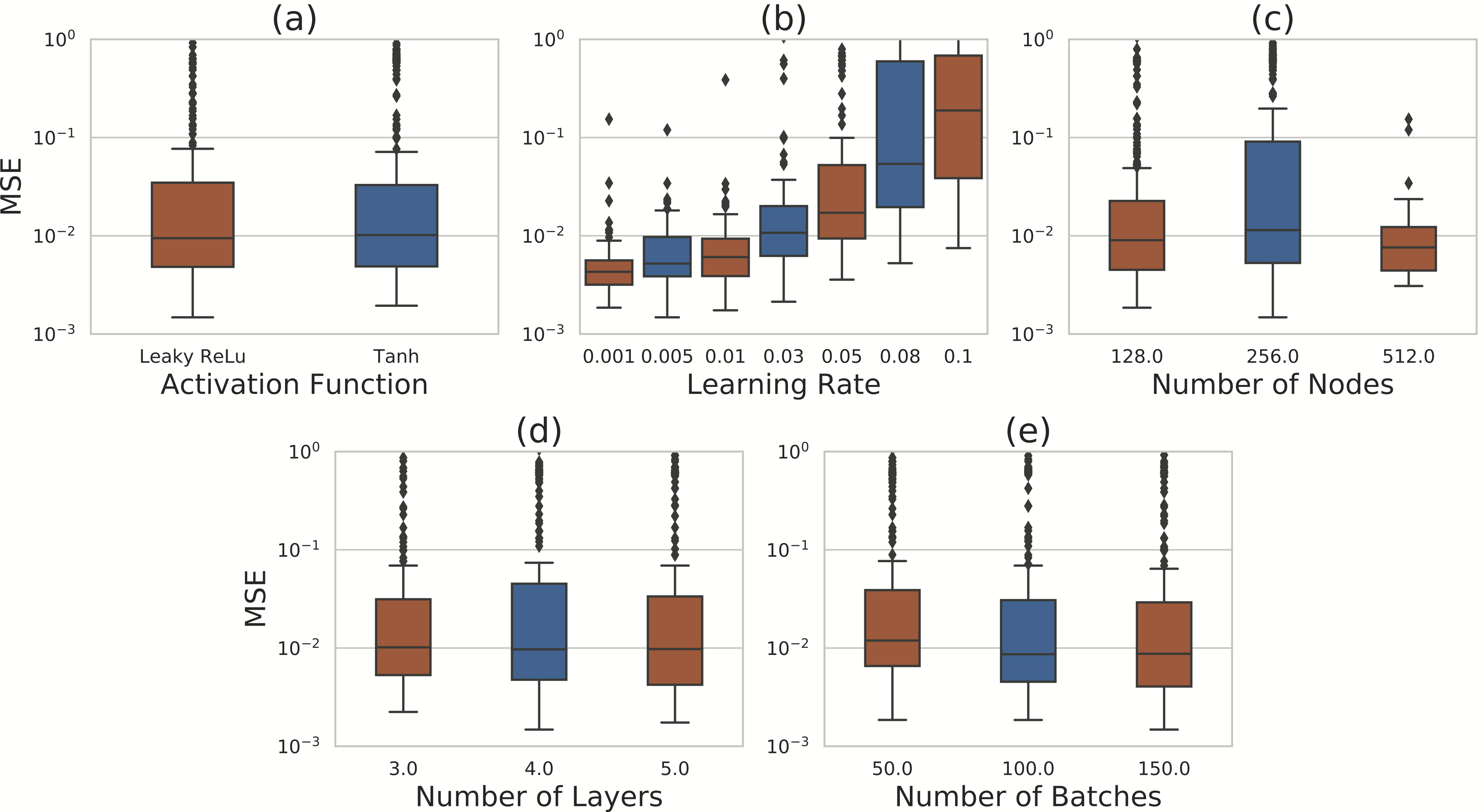}
	\caption{Hyper-parameter optimization used to determine a suitable architecture for the ANN. We swept through various parameters like the activation function (a), the optimizer's learning rate (b), the number of neurons (c), the number of layers (d), and the number of batches per epoch (e). Each box and whisker plot illustrates the distribution of a particular parameter with reference to its MSE after the final epoch. While some parameters showed little influence (number of layers, activation functions, etc) others greatly affected the MSE convergence (learning rate).}
	\label{fig:HPO}
\end{figure}

To identify a suitable ANN architecture, we performed a hyper-parameter optimization (HPO), where several ANNs with different architectures were simulated simultaneously. We swept through common ANN architecture components, like the number of layers, the number of neurons for each layer, each neuron's activation function, and the batch size. We concurrently trained 1200 different ANNs on 1200 cores using Brigham Young University's Fulton Supercomputing Lab and the TensorFlow package \cite{martin_abadi_tensorflow:_2015}. Each simulation took approximately 12 hours. Figure \ref{fig:HPO} illustrates the HPO's results. We measured the accuracy and effectiveness of each network by tracking the mean squared error (MSE) and coefficient of determination ($R^2$)  for all simulated permutations.\par

From the HPO, we chose to train an ANN with 8 layers. Each layer had 32, 64, 128, 256, 128, 64, 32, and 16 neurons respectively. Each neuron used a leaky ReLu activation function. No dropout was used.

\section{Device fabrication, measurement, and calibration}

We designed 11 different ICBGs each with a linear chirp of 6 nm. We designed 5 of the devices with a reversed chirping, such that their resultant group delay profiles would be mirror images of their counterparts. Some devices were 750 grating periods long and the others were 250. We chose corrugation widths of 10 nm, 30 nm, and 50 nm. \par

\begin{figure}
	\centering
	\includegraphics{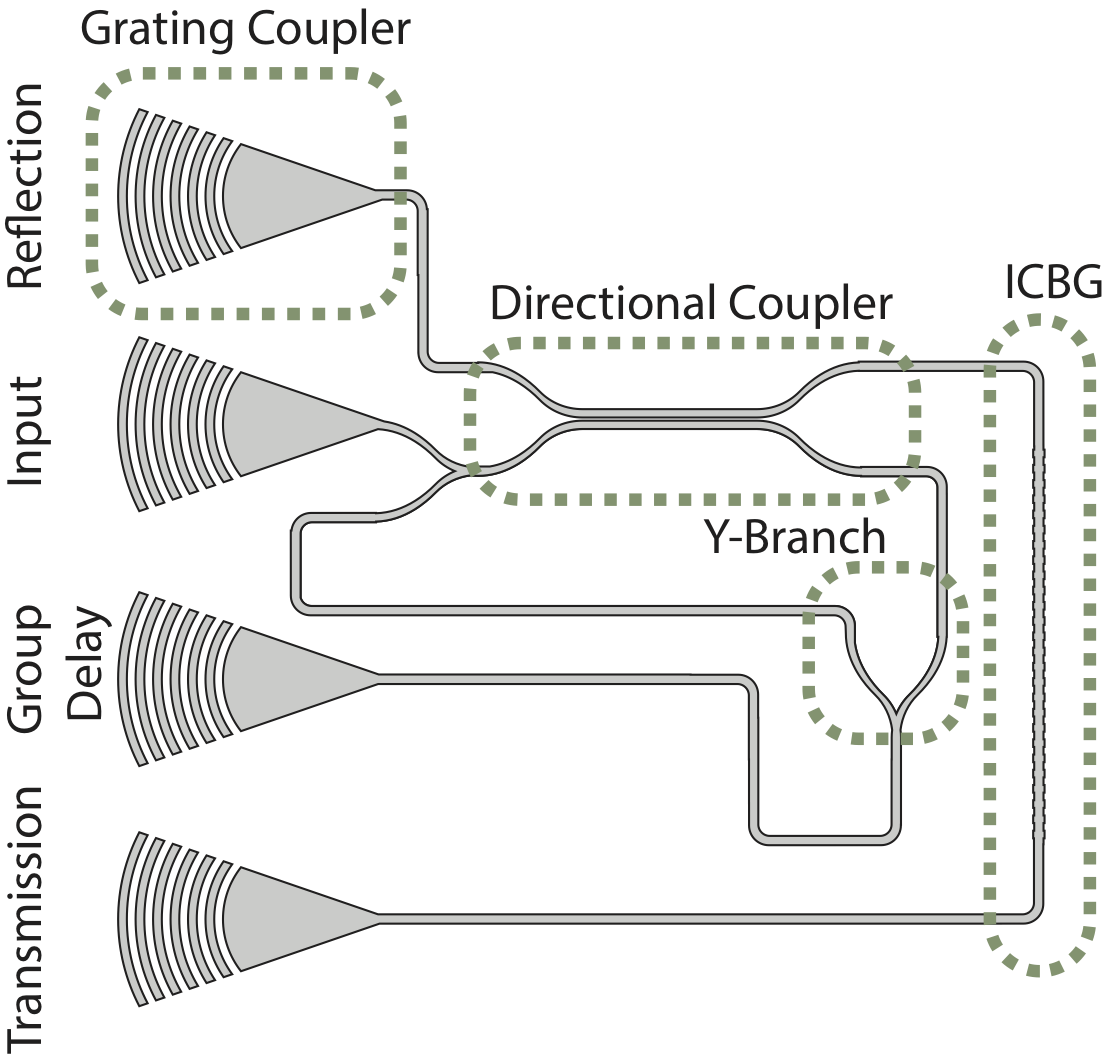}
	\caption{Interrogation circuit used to extract the reflection, transmission, and group delay profiles of a single ICBG simaltaneously. Light is routed on and off the chip using grating couplers. The group delay is extracted using a Mach-Zehnder Interferometer (MZI)  formed by various directional couplers and Y-branches.}
	\label{fig:fabrication}
\end{figure}

To efficiently extract the reflection, transmission, and group delay profiles of the same ICBG, we designed an interrogator circuit using various Y-branches, directional couplers, and grating couplers. Figure \ref{fig:fabrication} illustrates the circuit. We used the grating couplers to direct light on and off of the chip. We routed the light using the Y-branches and directional couplers. We interfered the reflection signal with the original reference signal using a Mach-Zhender Interferometer (MZI) in order to measure the group delay information. \par

Our devices were fabricated at the University of Washington in collaboration with the University of British Colombia and the SiEPIC program on a 150 mm silicon-on-insulator (SOI) wafer with 220 nm thick silicon on 3 $\mu$m thick silicon dioxide and a hydrogen silsesquioxane resist (HSQ, Dow-Corning XP-1541-006). Electron beam lithography was performed using a JEOL JBX-6300FS system operated at 100 keV energy \cite{bojko2011electron}, 8 nA beam current, and 500 $\mu$m exposure field size. The silicon was removed from unexposed areas using inductively coupled plasma etching in an Oxford Plasmalab System 100. Cladding oxide was deposited using plasma enhanced chemical vapor deposition (PECVD) in an Oxford Plasmalab System 100. \par

To characterize the devices, a custom-built automated test setup \cite{chrostowski_silicon_2015} with automated control software written in Python was used.  An Agilent 81600B tunable laser was used as the input source and Agilent 81635A optical power sensors as the output detectors. The wavelength was swept from 1500 to 1600 nm in 10 pm steps.  A polarization maintaining (PM) fibre was used to maintain the polarization state of the light, to couple the TE polarization into the grating couplers \cite{wang2014focusing}. A polarization maintaining fibre array fabricated by PLC Connections (Columbus OH, USA) was used to couple light in/out of the chip.

\begin{figure}
	\centering
	\includegraphics{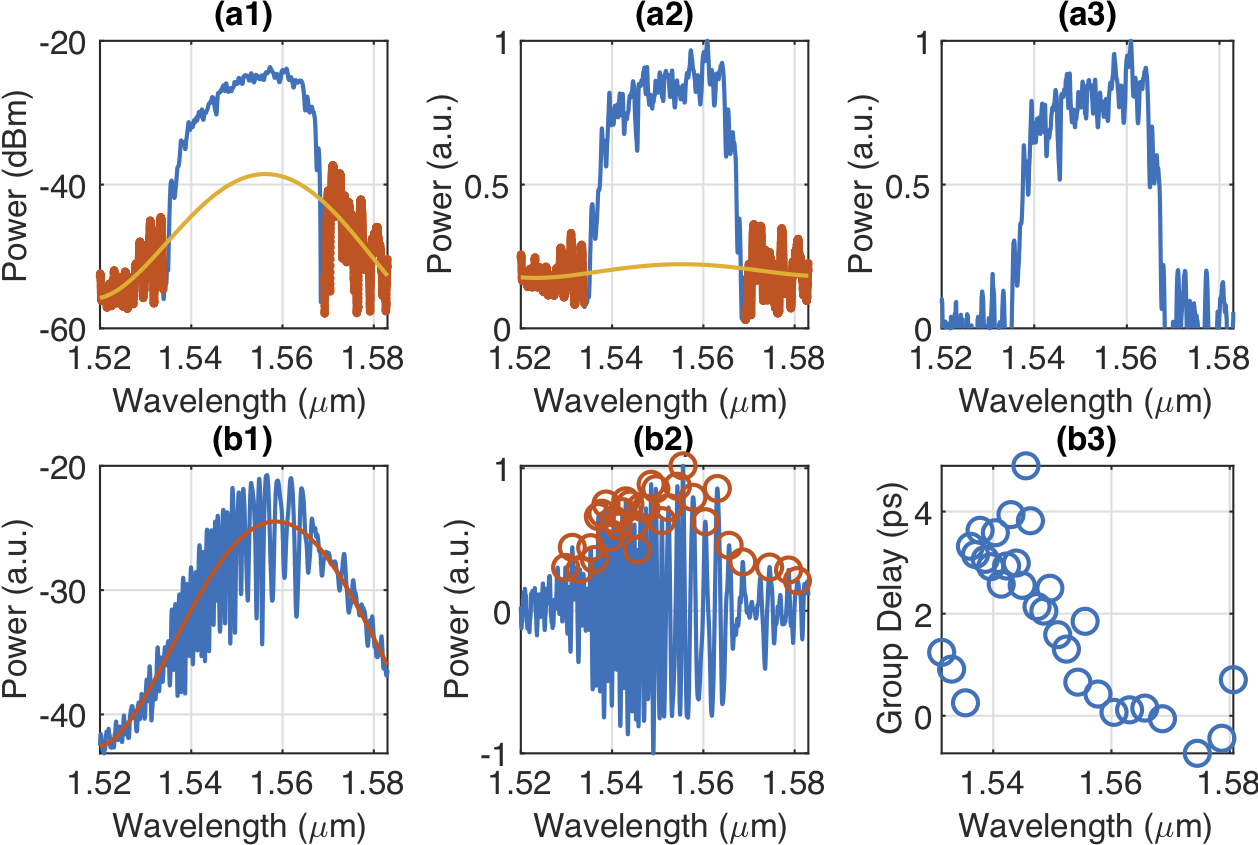}
	\caption{Calibration process used to extract the measured reflection and group delay responses. The reflection data is first fit to a fourth order polynomial outside of the expected bandwidth in order to remove the grating couplers' transfer function (a1). Next, the data is once again fit to a fourth order polynomial outside of the device's bandwidth to identify the noise floor (a2). The data is then normalized to unit power (a3). Similar to the reflection data, the group delay data is also fit to a fourth order polynomial to remove the grating couplers' response (b1). Next, the FSR is approximated using a peaktracking algorithm (b2). From the FSR, the group delay is estimated (b3).}
	\label{fig:calibration}
\end{figure}

To estimate the reflection and group delay profiles from the measurement data, we calibrated out the band-limited spectral responses induced by the grating couplers, directional couplers, and Y-branches. Figure \ref{fig:calibration} illustrates this process for both the reflection and group delay data. For the reflection measurements, we first fit the data outside of the ICBG's bandwidth to a fourth order polynomial. We use this polynomial fit to remove the couplers' responses. We then relocate the noise floor by fitting, once again, the data outside of the ICBG's bandwidth. To extract the group delay, we fit the entire MZI interference pattern to a fourth order polynomial to remove the couplers' response. We then estimate the free spectral range (FSR) of the interferometer using a peak-tracking algorithm. From the FSR, along with the relative path length difference ($L_ref$) of approximately 200 $\mu$m, we can estimate the group delay ($\tau$) using
\begin{equation}
\tau(\lambda)=\frac{(L_{ref}-L(\lambda))\cdot n_g(\lambda)}{c}
\end{equation}
where
\begin{equation}
L(\lambda)=\frac{\lambda^2}{FSR \cdot n_g(\lambda)}
\end{equation}
and $n_g(\lambda)$ is the group index of the reference arm waveguide.

\section{Experimental Results}

\begin{figure}
	\centering
	\includegraphics[width=5in]{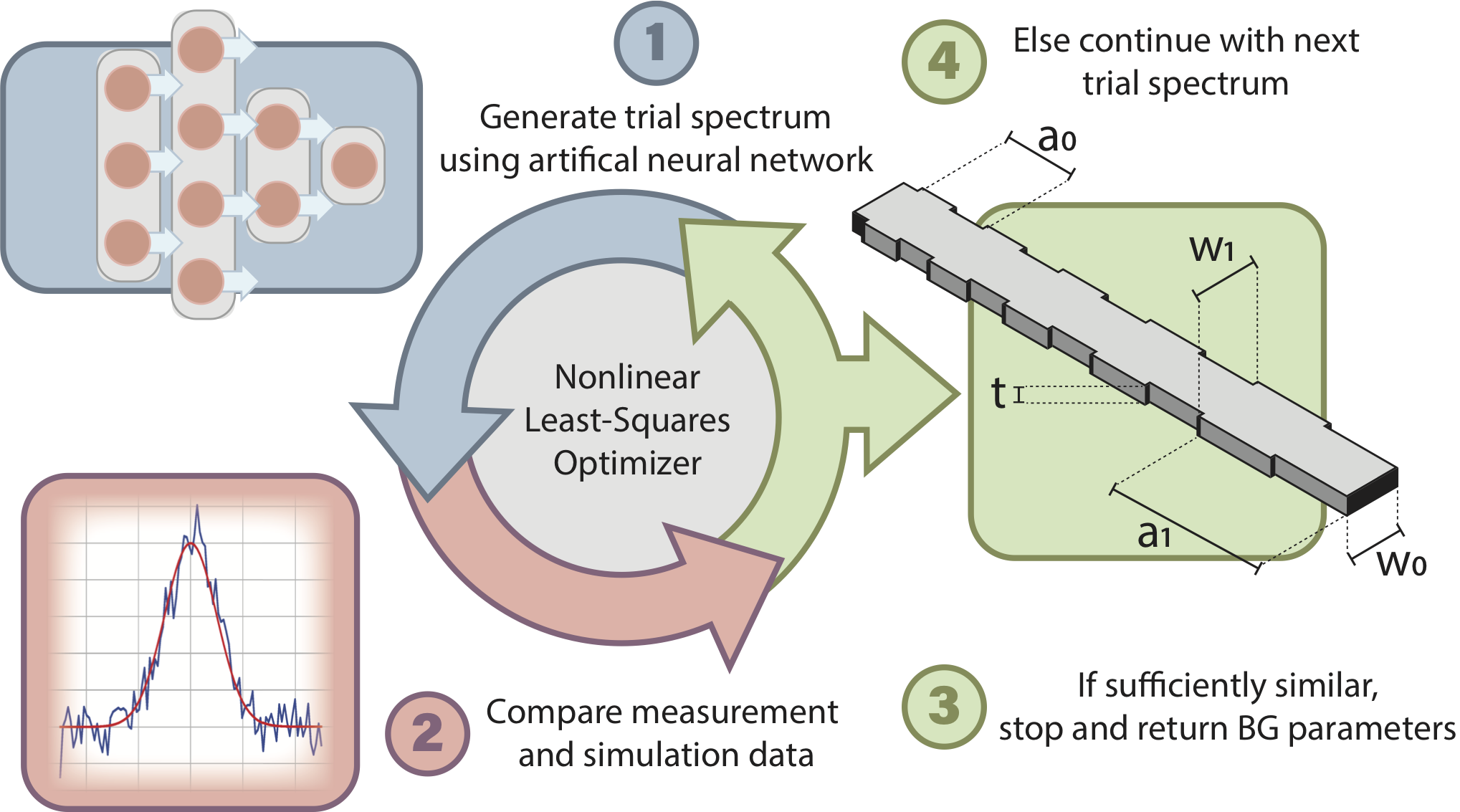}
	\caption{Efficient and robust method to extract fabricated ICBG device parameters using ANNs and a nonlinear least-squares optimizer. First, the ANN simulates reflection and group delay spectra for the device's initial design parameters (1). Then, the simulations are compared directly to the measured data (2). If the results are sufficiently similar, the optimizer returns the device parameters (3). If not, the optimizer strategically simulates a new set of device parameters based on the residual error (4).}
	\label{fig:processFlow}
\end{figure}

To estimate the actual fabrication parameters of the ICBGs, we used the ANN in conjunction with a nonlinear least squares fitting routine within the SciPy package \cite{scipy}. The routine initializes by calling the ANN using the original design parameters. The simulated reflection and group delay profiles are directly compared to the measurement data. From the residuals, the algorithm decides whether the current design is sufficiently similar to the measurement data or if further simulation is needed. Figure \ref{fig:processFlow} illustrates this procedure.

 Since the ICBG has fabrication limits that can be cast as parameter bounds, we chose to run a Trust Region Reflective (TRF) optimization algorithm  within the nonlinear solver \cite{branch1999subspace}. Specifically, we bounded the first and last ICBG periods ($a_0$ \& $a_1$) between 312 nm and 328 nm and the corrugation width between 1 nm and 50 nm. The number of periods was fixed.

\begin{figure}
	\centering
	\includegraphics{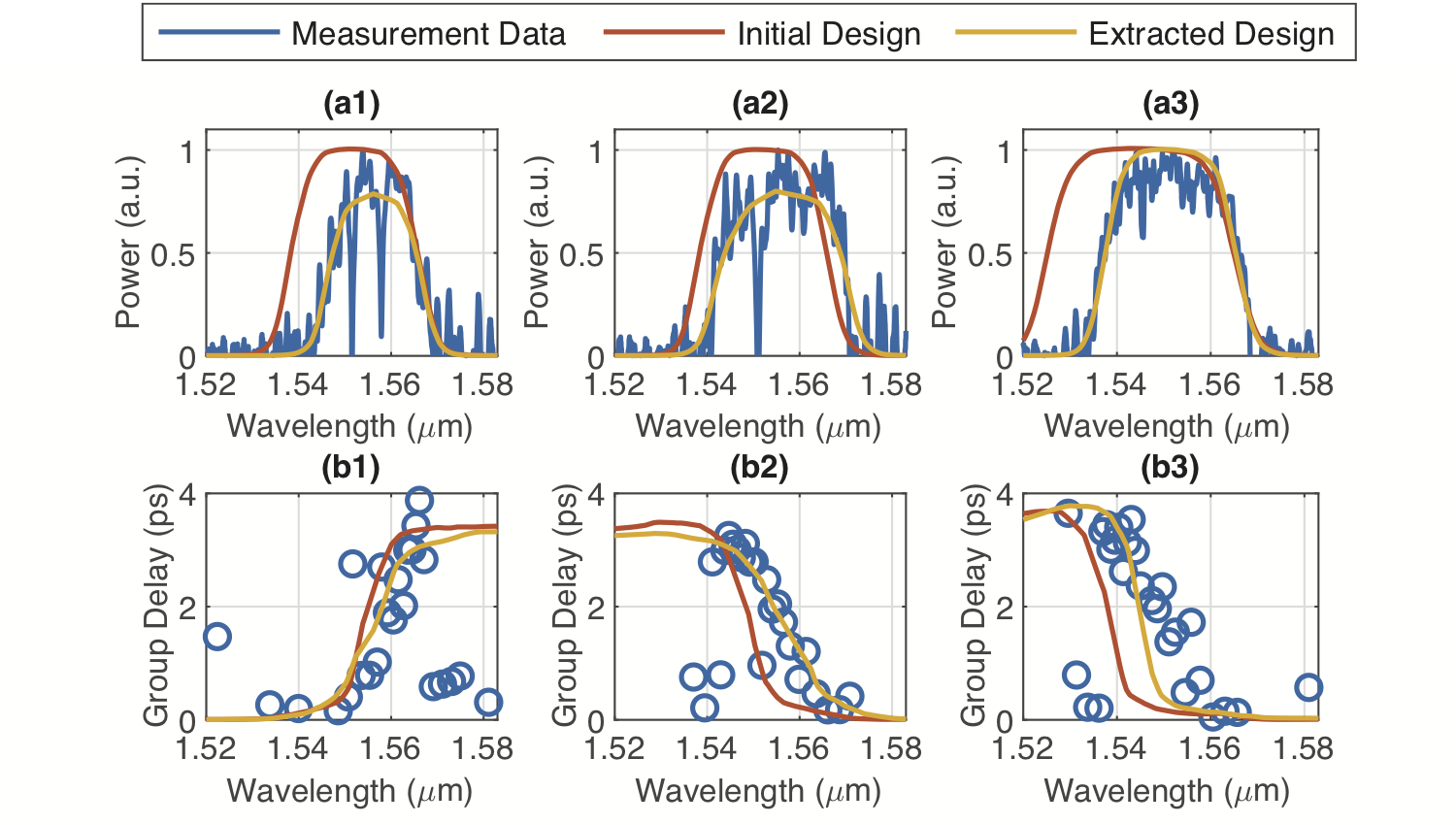}
	\caption{The extracted reflection (a1, a2, a3) and group delay (b1, b2, b3) profiles (yellow) compared to the initial design profiles (red) and the calibrated measurement data (blue).}
	\label{fig:fit}
\end{figure}

We chose to extract parameters of three different ICBGs. After just 5 minutes of optimization on a Macbook Air 2012 (1.8 GHz Intel Core i5, 4 GB 1600 MHz DDR3 RAM), the solver converged on new parameters for all three devices that more reasonably reflect the measurement data. Figure \ref{fig:fit} illustrates the algorithm's results compared to the fabrication data and the original design spectra. Table \ref{table:results}

\begin{table}[h!]
	\begin{center}
		\begin{tabular}{|c || c c | c c | c c|} 
			\hline
			& \multicolumn{2}{c|}{Device 1} & \multicolumn{2}{c}{Device 2} & \multicolumn{2}{|c|}{Device 3} \\ [0.5ex] 
			\hline
			Design Parameters & Design & Extracted & Design & Extracted & Design & Extracted \\
			\hline\hline
			$a_0$ (nm) & 324 & 324.3 & 318 & 315.8 & 318 & 320.1 \\  
			\hline
			$a_1$ (nm) & 318 & 317.0 & 324 & 325.6 & 324 & 323.7 \\
			\hline
			$NG$ & 750 & 750 & 750 & 750 & 750 & 750\\
			\hline
			$\Delta w$ (nm) & 30 & 13.7   & 30 & 15.7 & 50 & 39.3 \\
			\hline
		\end{tabular}
	\end{center}
	\caption{Parameter extraction results for three separate devices. The design parameters are compared directly to the algorithm's extracted parameters for each device..}
	\label{table:results}
\end{table}

Not only do the algorithm's profiles match the data much better, but the extracted parameter differences are expected from the processes used to fabricate the devices. For example, the algorithm predicts a slightly wider chirping bandwidth and smaller corrugation width for all three devices. The E-beam raster grid's resolution approaches the chirping resolution of the ICBG (1 nm), so "snapping" from one grid point to the next results in slightly wider chirping bandwidths. The E-beam's resolution, along with the etch process, also tend to round the sharp ICBG corners, resulting in lower net corrugation width.

Other small differences between the extracted parameter sets and the fabricated data, like the fabry-perot resonances, are difficult to model with the current ANN abstraction. It would require a much more sophisticated, and possibly impractical, parameterization to capture these defects. Despite these small discrepancies, the fitting algorithm and ANN demonstrate a strong ability to extract parameters for complex silicon photonic devices.

\section{Conclusion}

We demonstrate a novel silicon photonic parameter extraction method using artificial neural networks. Our method is capable of extracting parameters for complicated devices, like integrated chirped Bragg gratings, without sacrificing the speed of traditional analytic methods. To validate our method, we fabricated and measured various integrated chirped Bragg gratings and extracted the actual parameters. Future work will explore other parameterizations and new devices, like adjustable splitters and directional couplers.

\section*{Acknowledgments}
We acknowledge the edX UBCx Phot1x Silicon Photonics Design, Fabrication and Data Analysis course, which is supported by the Natural Sciences and Engineering Research Council of Canada (NSERC) Silicon Electronic-Photonic Integrated Circuits (SiEPIC) Program. The devices were fabricated by Richard Bojko at the University of Washington Washington Nanofabrication Facility, part of the National Science Foundation’s National Nanotechnology Infrastructure Network (NNIN). Enxiao Luan performed the measurements at The University of British Columbia.

\section*{Disclosures}
The authors declare that there are no conflicts of interest related to this article.


\bibliography{main}






\end{document}